\documentclass[10pt]{iopart}

\usepackage[latin1]{inputenc}
\usepackage{color}
\usepackage{xspace}
\usepackage{graphicx}
\usepackage{iopams}
\usepackage{xspace}
\usepackage{dcolumn}
\usepackage{bm}
\usepackage{color}
\usepackage{citesort}
\usepackage{stfloats}
\usepackage[extra]{tipa}
\usepackage{times}

\newcommand{\onlinecite}{\cite}
\newfont{\tensy}{cmsy10}

\ProvideTextCommand{\DJ}{OT1}{\raisebox{0.25ex}{-}\kern-0.3em D}

\begin{document}

\topical[]{Phase diagram of interacting spinless fermions on the honeycomb lattice}

\author{Sylvain Capponi}
\address{Laboratoire de Physique Th\'eorique, Universit\'e de Toulouse, CNRS, UPS, France}
\ead{capponi@irsamc.ups-tlse.fr}

\begin{abstract}
Fermions hopping on a hexagonal lattice represent one of the most active research field in condensed matter since the discovery of graphene in 2004 and its numerous applications. Another exciting aspect of the interplay between geometry and quantum mechanical effects is given by the Haldane model (F. D. M. Haldane, Phys. Rev. Lett. {\bf 61}, 2015 (1988)) where spinless fermions experiencing a certain flux pattern on the honeycomb  lattice leads to the stabilization of a topological phase of matter, distinct from a Mott insulator and dubbed Chern insulator nowadays. 
In this context, it is crucial to understand the role of interactions and this review will describe recent results that have been obtained for a minimal model, namely spinless fermions fermions with nearest and next-nearest neighbour
 density-density interactions on the honeycomb lattice at half-filling.
 
 Topics addressed include an introduction of the minimal model and a discussion of the possible instabilities of the Dirac semimetal, a presentation of various theroretical and numerical approaches, and a summary of the results with a particular emphasis on the stability or not of some exotic quantum phases such as charge ordered ones (similar to Wigner crystals) or spontaneous Chern insulator phase.
\end{abstract}

\tableofcontents

\setlength\hoffset{-0.5in}\setlength\voffset{-0.5in}\setlength\textwidth{6.75in}
\setlength\columnsep{0.2in}\setlength\textheight{9.25in}\mathindent=0.in\twocolumn

\section{Introduction}\label{sec:intro}

Strongly correlated fermionic systems are exciting because they can host a variety of unconventional exotic quantum phases of matter, hence possessing very rich phase diagrams~\cite{Dagotto2005}. 
On the other hand, even weakly- or non-interacting fermions can exhibit several phases including trivial ones (insulator, semi-metal, metal), but also topological phases of matter, i.e. phases with cannot be adiabatically connected to trivial insulators for instance, such as the quantum Hall effect~\cite{book_QHE} or the topological insulators in the presence of strong spin-orbit coupling~\cite{topo_review2,topo_review}. Key signatures of these topological phases are the existence of protected (charge or spin) edge states. Some effects of correlations on such topological phases have already been reviewed in Ref.~\onlinecite{Hohenadler2013}.

In a seminal paper~\cite{Haldane1988}, Haldane has shown how spinless fermions hopping on a honeycomb lattice with a particular flux pattern (but no net flux per plaquette) can realize a topological insulating phase, dubbed nowadays a Chern insulator. This is a way to realize a quantum Hall state without magnetic field and has generated a lot of activity. On the experimental side, a great motivation came from the discovery of graphene~\cite{Novoselov2004}, which is a purely two-dimensional material consisting of a one-atom-thick sheet of graphite. 
In 2005, Kane and Mele have proposed that graphene could realize a spin topological insulator with helical edge states provided its intrinsic spin-orbit coupling would be large enough~\cite{Kane2005}. Unfortunately, {\it ab initio} calculations~\cite{Min2006} point to a rather small spin-orbit coupling (of order $0.01$~K). Nevertheless, in spite of this negative results, there could be other ways to realize a topological phase, for instance using longer-range (or Coulomb) interactions as we will discuss in details below. Although this is an appealing roadmap to follow in order to stabilize some exotic phases, this remains quite challenging since strongly correlated systems cannot often be studied in some unbiased way. Our main interest in this review is to understand whether interactions \emph{alone} can generate a topological phase starting from a \emph{trivial} non-interacting phase. For the sake of the argument, we will consider only the simplest situation 

\section{Minimal model for interacting spinless fermions on the honeycomb lattice}

\begin{figure}[!ht]
\centering
\includegraphics[width=0.85\linewidth]{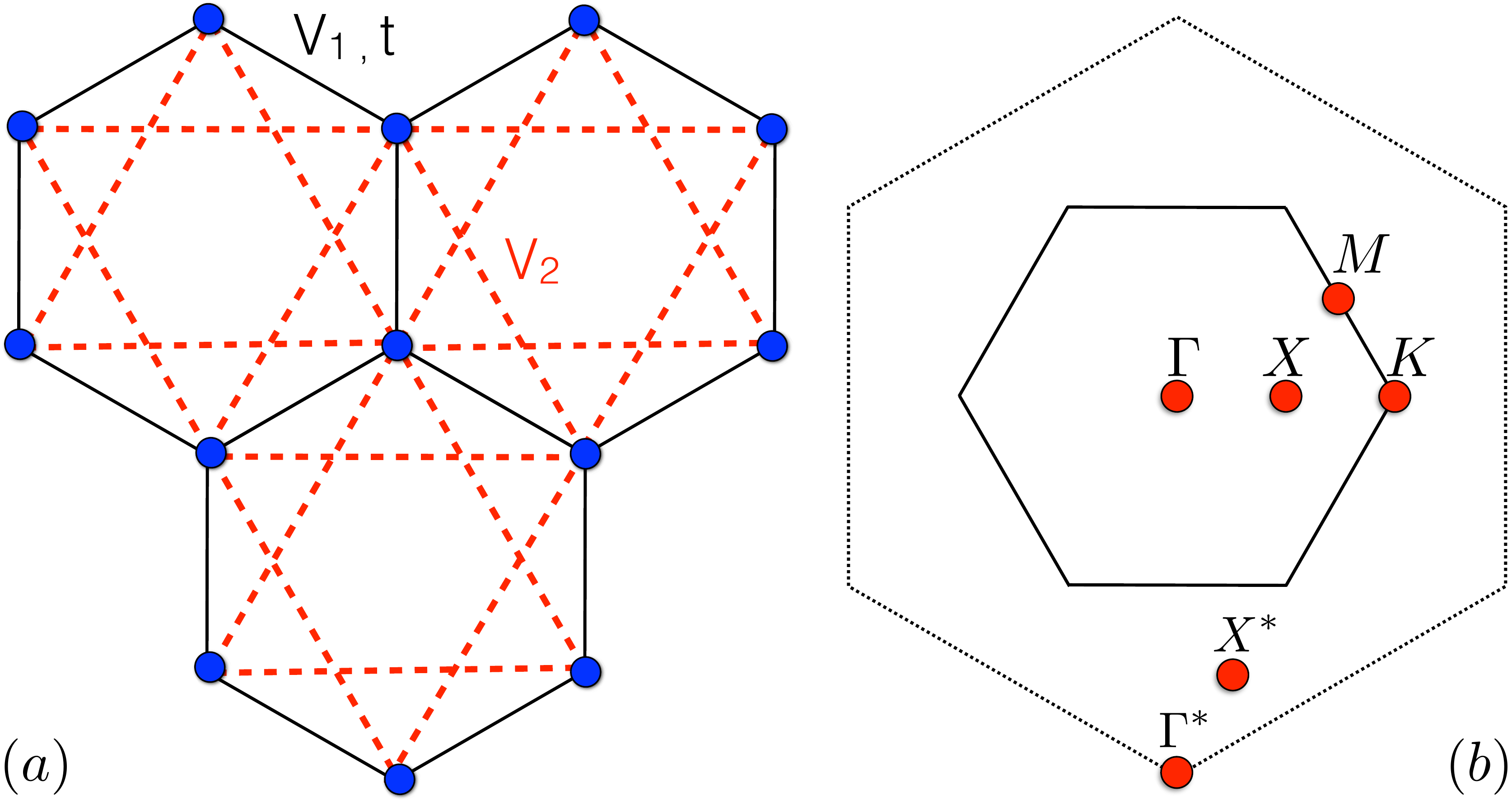}
\caption{(Color online) 
(a) Illustration of the honeycomb lattice with $V_1$, $V_2$ interactions and the hopping $t$.
(b) First (solid line) and second (dashed line) Brillouin zone of the honeycomb lattice including the 
 location of a few special points in the Brillouin zone.
}
\label{fig:latticeBZ}
\end{figure}

In order to focus on the simplest relevant model for the physics that we are discussing, we will consider interacting spinless fermions at half-filling on a 
 honeycomb lattice, i.e. the following Hamiltonian~:
\begin{eqnarray}
\label{eq:Hspinless}
{\cal H} & = & -t\sum_{\langle i j\rangle} (c^\dagger_i c_j +h.c.) + V_1 \sum_{\langle i j\rangle} (n_i -1/2)(n_j-1/2) \nonumber\\
&+&V_2 \sum_{\langle\langle i j\rangle\rangle} (n_i -1/2)( n_j-1/2)
\end{eqnarray}
depicted in Fig.~\ref{fig:latticeBZ}(a), 
where $c_i$ and $c^\dagger_i$ are the spinless fermionic operators, $t=1$ is the nearest-neighbor hopping amplitude, $V_1$ and $V_2$ are the density-density repulsion or attraction strengths respectively on nearest- (NN) and next-nearest neighbors (NNN). 

For completeness, we will also discuss some results obtained on its spinful extension:
\begin{eqnarray}
\label{eq:Hspinfull}
{\cal H} & = & -t\sum_{\langle i j\rangle, \, \sigma} (c^\dagger_{i\sigma} c_{j\sigma} +h.c.) 
+ U \sum_i n_{i\uparrow} n_{i\downarrow}\\
&+& V_1 \sum_{\langle i j\rangle} (n_i -1)(n_j-1) + V_2 \sum_{\langle\langle i j\rangle\rangle} (n_i -1)( n_j-1) \nonumber 
\end{eqnarray}
where the fermions carry a spin 1/2 ($\sigma=\uparrow,\, \downarrow$) and $U$ corresponds to the onsite Hubbard interaction.

\section{Theoretical approaches and phase diagrams}

In this section, we will review various analytical and numerical approaches that have been applied to model (\ref{eq:Hspinless}). 
We will describe the different possible phases that have been proposed, and show when available the corresponding phase diagrams. 

\subsection{Instabilities of the Dirac semi-metal}

Let us start by considering the non-interacting spinless case. From the seminal paper by Wallace~\cite{Wallace1947}, it is known that, at half-filling, the band structure exhibits two Dirac cones in the Brillouin zone. 
This is theoretically very appealing since it allows to observe relativistic phenomena in condensed matter systems~\cite{Goerbig2013} and has been experimentally realized first with the discovery of graphene~\cite{Novoselov2004}. 

Now, when considering the two-dimensional Dirac equation, it is possible to classify the relevant perturbations that could open a gap:
\begin{itemize}
\item If one adds a potential term that breaks the sublattice symmetry but preserves time-reversal symmetry, then a 
N\'eel CDW I can emerge~\cite{Semenoff1984}.
\item If one modulates the nearest-neighbor hopping amplitudes then a 
Kekul\'e bond-density wave (that preserves sublattice and time-reversal symmetries) emerges~\cite{Hou2007}.
\item In the presence of next-nearest-neighbor hopping with fluxes (but no net total flux per hexagon), then the 
QAH bond-density wave (that breaks both sublattice and time-reversal symmetries) can be stabilized as shown by Haldane~\cite{Haldane1988}.
\end{itemize}
These phases can thus be understood as resulting from the relevant effects of 4 different mass terms~\cite{Ryu2009,Herbut2009}. 

For comparison, the spinful case is far much richer with 36 masses that can be added to the Dirac equation so that several different phases can be realized~\cite{Ryu2009}. 

Finally, let us remind the reader that due to a vanishing density of states in the noninteracting semi-metallic phase, a finite $V_1$ and/or $V_2$ is needed for any kind of instability. 

\subsection{Mean-field analyses} 

\begin{figure}[!hb]
\includegraphics[width=0.88\linewidth]{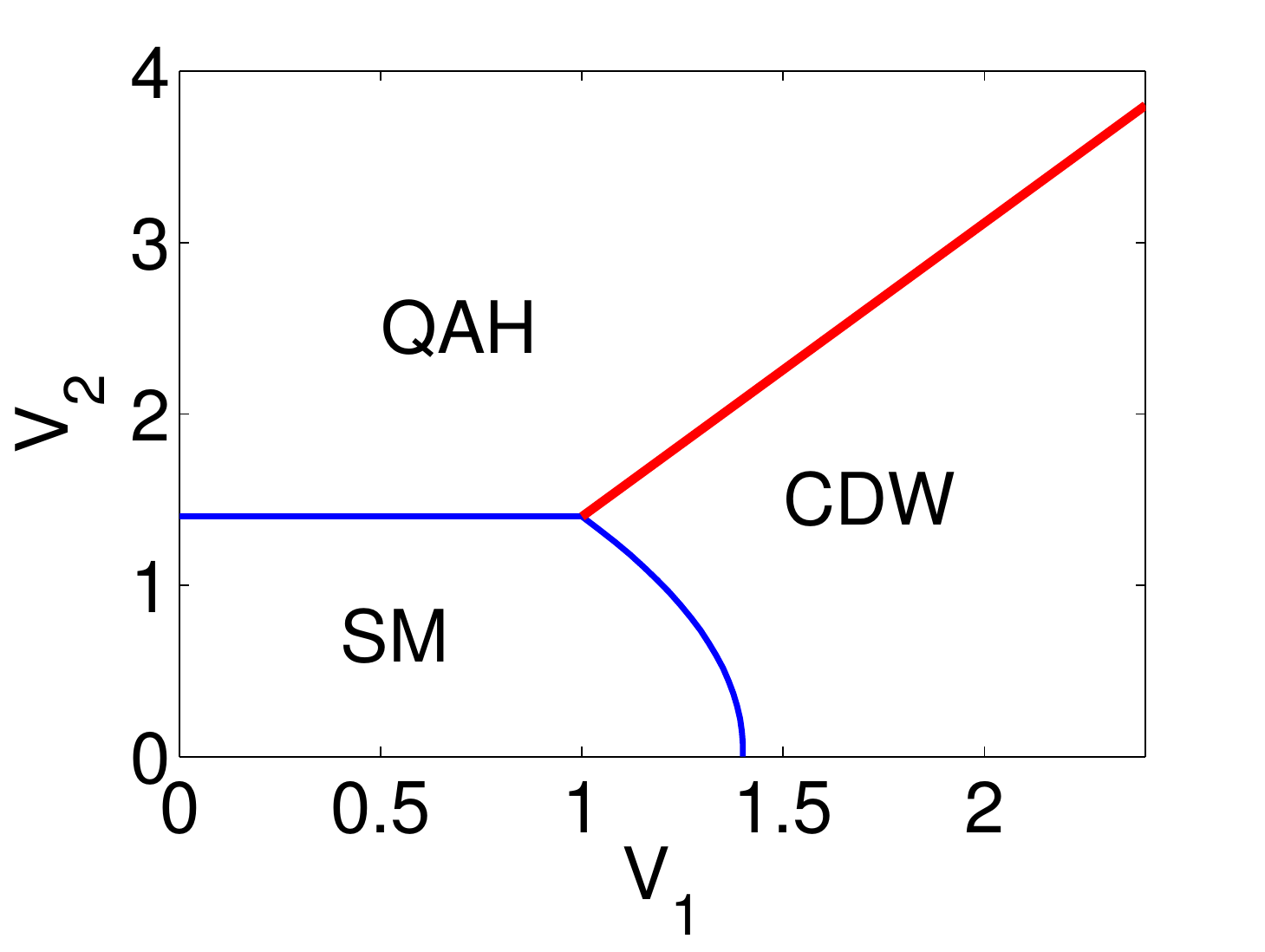}
\caption{(Color online) Phase diagram for model (\ref{eq:Hspinless}) vs $(V_1/t,V_2/t)$. The semimetallic (SM) state that occurs at weak coupling is separated from the CDW and the topological QAH states via a continuous transition. The line separating the QAH and CDW marks a first-order transition, which terminates at a bicritical point.
Reprinted figure with permission from Ref.~\onlinecite{Raghu2008} Copyright (2008) by the American Physical Society.
}
\label{fig:MF1}
\end{figure}

Given all the possible instabilities of the semi-metallic phase, first attempts were made to investigate the zero-temperature phase diagram using mean-field analysis. 
In a seminal paper, Raghu {\it et al.}~\cite{Raghu2008} have solved the mean-field equations using the smallest unit cell. Their phase diagram is shown in Fig.~\ref{fig:MF1}. Besides the expected N\'eel CDW at large $V_1>0$, they have found the emergence of a large quantum anomalous Hall (QAH) phase for large $V_2>0$. This phase can be characterized by the existence of spontaneous charge currents. It could also be potentially realized in strained graphene~\cite{Roy2013a}.

\begin{figure}[!hb]
\includegraphics[width=0.8\linewidth]{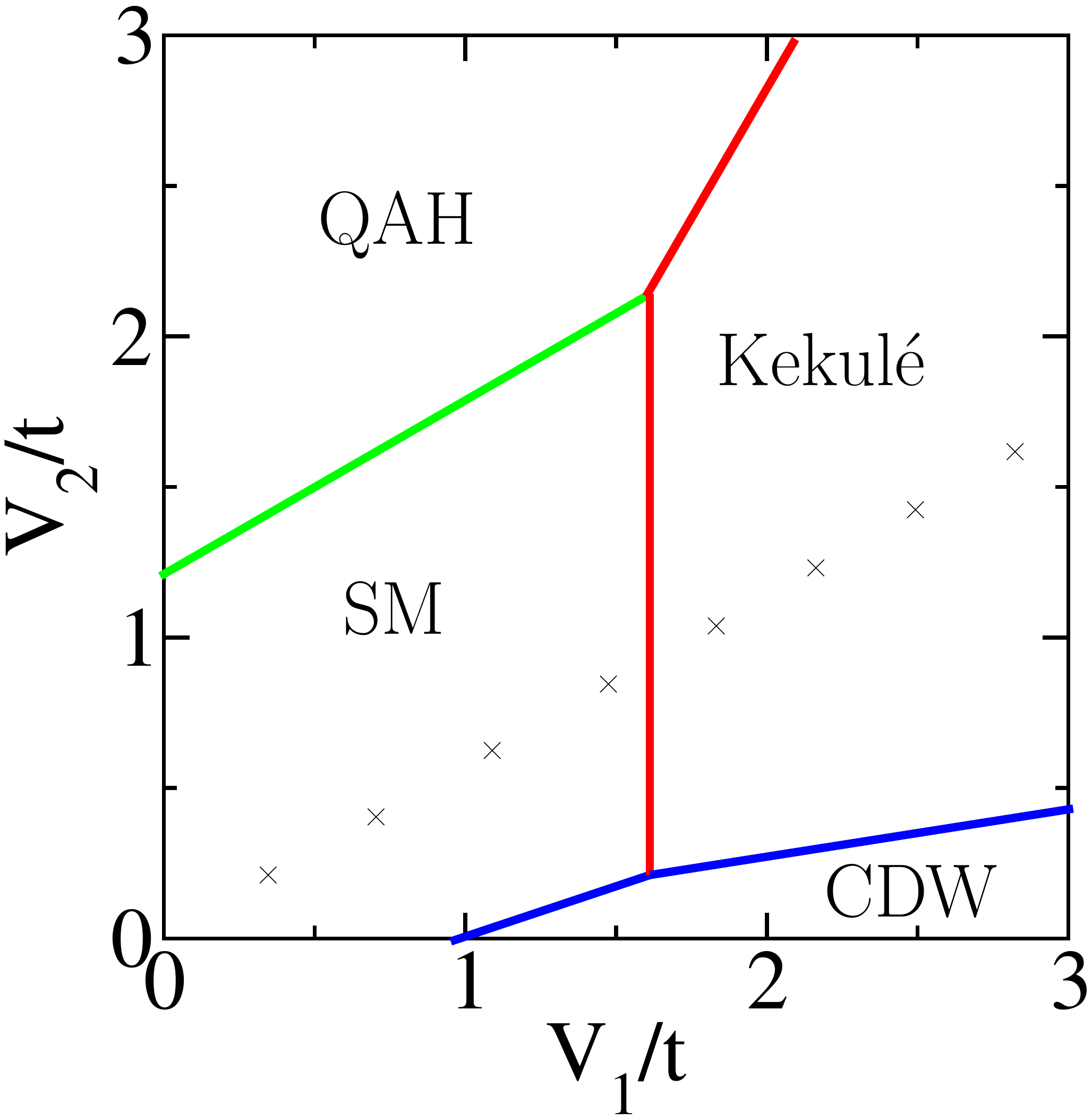}
\caption{(Color online) Same as Fig.~\ref{fig:MF1} using a more refined analysis. A novel Kekul\'e phase is proposed when both interactions are large enough. At the mean-field level all transitions from the SM phase are second order whereas transitions between all the gapped phases are first order. The  crosses represent the relevant line for  graphene based on a crude estimate of the bare Coulomb repulsion~\cite{Drut2009}.
Critical values along both axis are $V_1^c/t \simeq 0.93$ and $V_2^c/t\simeq 1.2$.
Reprinted figure with permission from Ref.~\onlinecite{Weeks2010} Copyright (2010) by the American Physical Society.
}
\label{fig:MF2}
\end{figure}

Allowing for additional order parameters, Weeks and Franz~\cite{Weeks2010} have obtained a slightly refined phase diagram (see Fig.~\ref{fig:MF2}) including a rather large Kekul\'e phase, which is indeed another potential candidate, see above. In particular, a rough estimate of the Coulomb interaction strength in graphene would suggest that the Kekul\'e phase could be realized experimentally. Unfortunately, suspended graphene samples (where interactions are stronger than for graphene over a substrate \cite{Du2008}) remain semi-metallic\cite{Du2009} down to low temperature ($\sim 1$~K). 

\begin{figure}[!ht]
\includegraphics[width=0.9\linewidth]{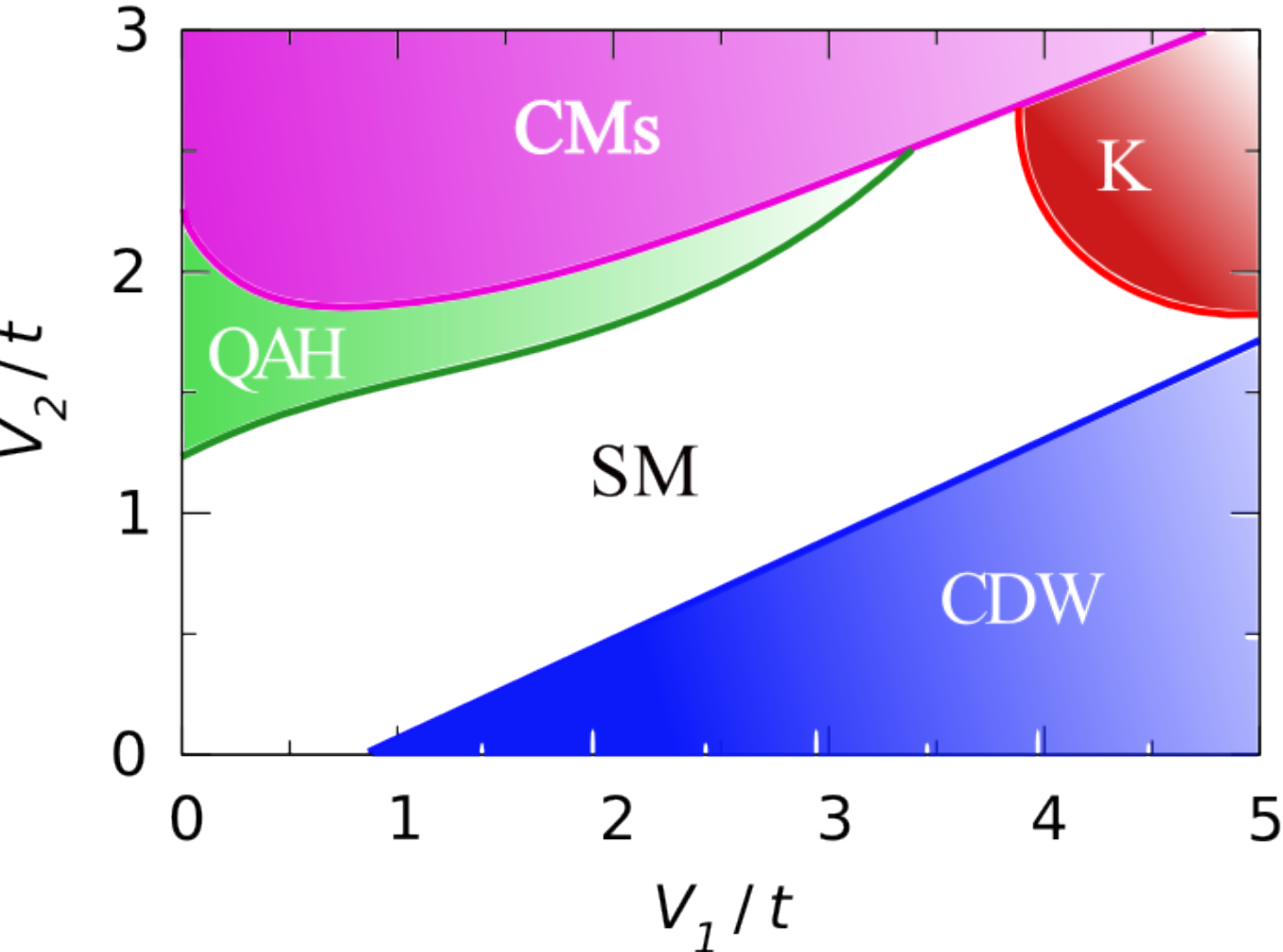}
\caption{(Color online) Same as Fig.~\ref{fig:MF2} using a larger unit cell. Kekul\'e phase is denoted with the letter K. Lines are guides to the eyes. CMs stands for a charge modulated phase which appears at large $V_2/t$.
Reprinted figure with permission from Ref.~\onlinecite{Grushin2013} Copyright (2013) by the American Physical Society.
}
\label{fig:MF3}
\end{figure}

Then, a larger (tripled) unit cell, allowing more instabilities, was used in Refs.~\onlinecite{Castro2011,Grushin2013}. There it was found that not only Kekul\'e phase can appear, but also the more interesting topological QAH can still be stabilized for a large range of parameters. However, in the latter study~\cite{Grushin2013},  the topological QAH phase  has shrunk substantially due to the
emergence of a charge-modulated (CM) phase (see Fig.~\ref{fig:MF3}) . This phase is characterized  by a larger unit cell and a distribution of charges on each hexagon as $(\rho,-\rho,\rho,-\rho,-\rho-\Delta,\rho+\Delta)$. Physically, the additional modulation allows to gain some energy by reducing the NNN repulsion $V_2$ at the cost of some unsatisfied $V_1$ bonds.

As a partial conclusion, we can observe that mean-field analysis has been used as an attempt to identify potential interesting phases in some microscopic models and to make connection with field theory predictions. However, being uncontrolled approximation, it can miss some phases and cannot locate accurately phase boundaries. Therefore, it should always be used with some caution. So we now turn to strong coupling analysis that can shed light from an opposite perspective.

\subsection{Strong coupling regime analysis}

In our recent work~\cite{Capponi2015}, we have suggested to investigate the strong-coupling regime, i.e. $|V_1/t|,\, |V_2/t| \gg 1$, starting from the analysis of the \emph{classical} ground-states.

Using a systematic enumeration on finite clusters, we have established the $t=0$ classical phase diagram shown in Fig.~\ref{fig:IsingPhaseDiagram}(a). For simplicity in the representation, we have used the equivalent notations using an angle $\theta \in [0,2\pi)$ so that $V_1=\cos\theta$ and $V_2=\sin\theta$. Without repeating all details that can be found in Ref.~\onlinecite{Capponi2015}, let us point out some remarkable features:

\begin{figure}[b]\setlength{\hfuzz}{1.1\columnwidth}
\begin{minipage}{\textwidth}
\centering
\includegraphics[width=0.82\linewidth,clip]{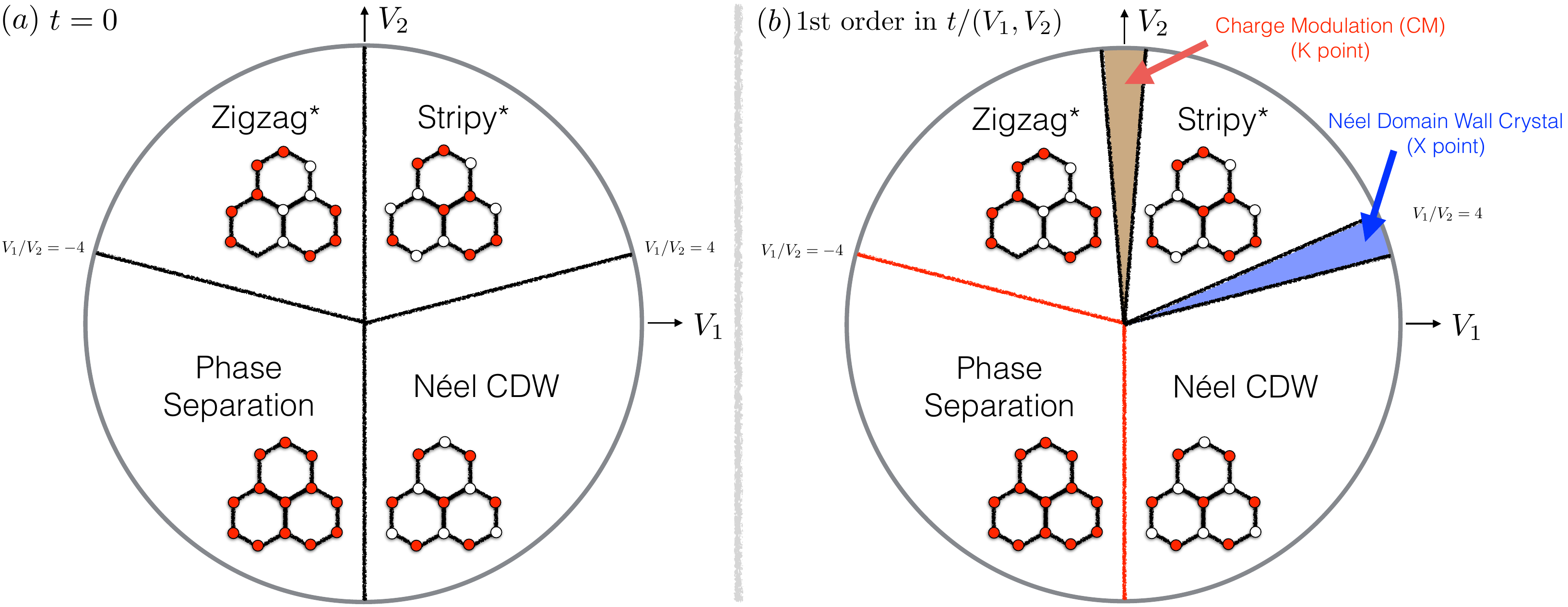}
\caption{(Color online) Classical phase diagram (a) and qualitative first order in $t/(V_1,V_2)$ phase diagram (b). The "Zigzag*" and the "Stripy*"  phases feature a
nontrivial ground state degeneracy (hence the "*" suffix in "Zigzag*" and "Stripy*"). The three points $V_2=1, V_1=\pm 4$ and $V_2=1, V_1=0$ feature an
extensive ground state degeneracy at the classical level. Upon including the first order correction due to the finite hopping $t$, two of these points spawn
new phases. The $V_2=1, V_1=0$ point develops into the charge modulation (CM) phase, while the $V_2=1, V_1=+4$ point broadens into 
a novel "N\'eel domain wall crystal" (NDWC), which is sketched in Fig.~\ref{fig:gs_eff_model_V1V2_4}. All regions and lines in (b) beyond the CM and NDWC phases
have no first order (in $t$) quantum corrections. 
Figure taken from Ref.~\onlinecite{Capponi2015}.
}
\label{fig:IsingPhaseDiagram}
\end{minipage}
\end{figure}

\noindent(i) For $\theta=\pm \pi/2$ (corresponding to Ising models on two decoupled triangular lattices), $\theta=\arctan(1/4)$, and $\theta=\pi-\arctan(1/4)$, there are an extensive number of ground-states; (ii) in the lower part of the phase diagram, we have found regions with two-fold degenerate ground-states, corresponding on one-side to standard N\'eel CDW phase (with particles occupying only one sublattice) and on the other side to a region of phase separation where the system prefers to be either  empty or completely filled with fermions; (iii) in the upper part, the number of ground-states increases with increasing system size $N$, and the patterns that we observe on both sides correspond respectively to zigzag or stripy patterns, possibly with defects.

Given the large degeneracy of the frustrated classical model in some large portions of the parameters space, one expects on general grounds that, in the presence of a finite hopping term $t$, quantum fluctuations will select some ordered state through an order-by-disorder mechanism. 
Thus, one possible attempt to detect this order consists in diagonalizing the kinetic operator \emph{projected onto the classical manifold}. Note that this is a major simplification compared to solving the full quantum mechanical problem.

 Performing this job, we have indeed found for instance that for $\pi/2<\theta<\pi-\arctan(1/4)$, there is a selection of a pristine six-fold degenerate zigzag state. Note that a similar state appears in the spin Heisenberg-Kitaev model on the honeycomb lattice~\cite{Chaloupka2013}.
 Unfortunately, we could not detect any selection among the stripy states on available sizes, although we expect some ordering, presumably with a large unit cell.

\begin{figure}[!htbp]
\includegraphics[width=\linewidth,clip]{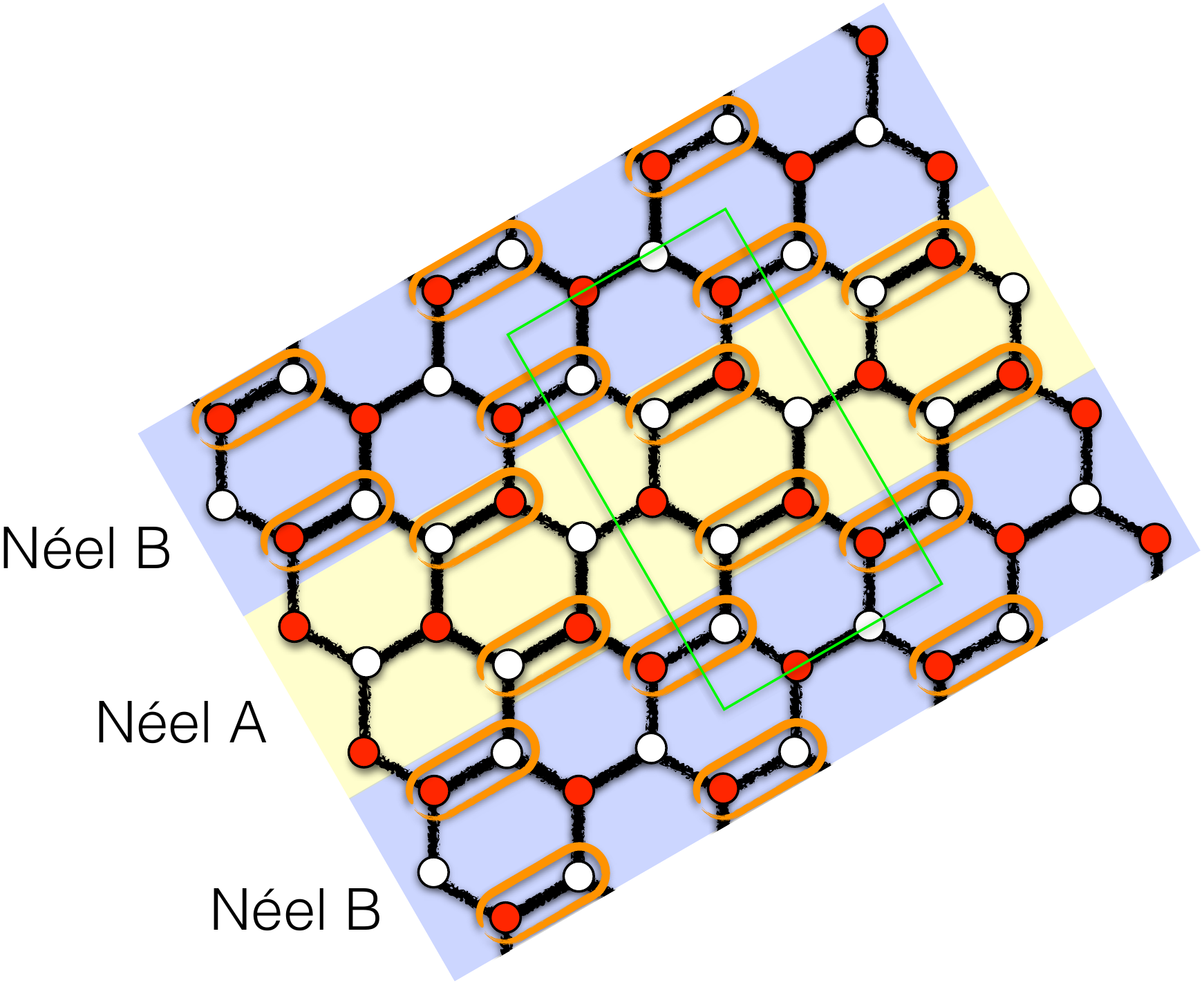}
\caption{(Color online) The N\'eel Domain Wall Crystal (NDWC): sketch of a classical ground state at $V_1=4,\ V_2=1$ on the $N=24$ sample, 
which is maximally flippable within the classical ground state manifold with respect to the hopping~$t$.
The shaded regions denote the two N\'eel domains, and the orange circled bonds along the domain walls
are flippable to first order in~$t$. The green box indicates a twelve-site unit cell.
Figure taken from Ref.~\onlinecite{Capponi2015}. 
}
\label{fig:gs_eff_model_V1V2_4}
\end{figure}

Then, let us discuss the selection in the massively degenerate regions. For $\theta=\pi/2$, the kinetic term selects the 18 maximally flippable states, which correspond to the charge modulated (CM) phase with a tripled unit cell and a sublattice imbalance. Note that this phase is an insulating one. For $\theta=\arctan(1/4)$, our finding was the selection of alternating strips of the two N\'eel CDW states in a particular arrangement shown in Fig.~\ref{fig:gs_eff_model_V1V2_4}, thus resulting in another 18-fold degenerate ground-state in the quantum case.

In summary of this part, the phase diagram in the strong coupling limit is shown in Fig.~\ref{fig:IsingPhaseDiagram}(b). 

\enlargethispage{-16.5\baselineskip}

\newpage

\subsection{Numerical approaches}

Based on the previous arguments, we expect a rather rich phase diagram, with many competing phases including a putative topological QAH one. We will now discuss  numerical investigations using state-of-the-art techniques for strongly correlated systems, namely Exact Diagonalization (ED), density-matrix Renormalization Group (DMRG),  quantum Monte-Carlo (QMC) and functional renormalization-group (fRG) technique.

\subsubsection{Exact Diagonalization}
-- Obviously, ED technique is called for as it is unbiased and could allow to discriminate between the different competing phases. The major caveat is of course size limitation which can prevent conclusion on the thermodynamic limit, or the difficulty to tackle with incommensurate phases for instance~\footnote{Note that dealing with incommensurate phases is also difficult for mean-field approaches.}. Therefore, we will argue that a systematic study should be performed, i.e. one should consider different cluster sizes (with different $k$ points in their Brillouin zone) that can accomodate various phases and also one should try to perform finite-size scaling (even though it is limited typically to less than 50 sites in the spinless case). 

\begin{figure}[!h]
\includegraphics[width=\linewidth]{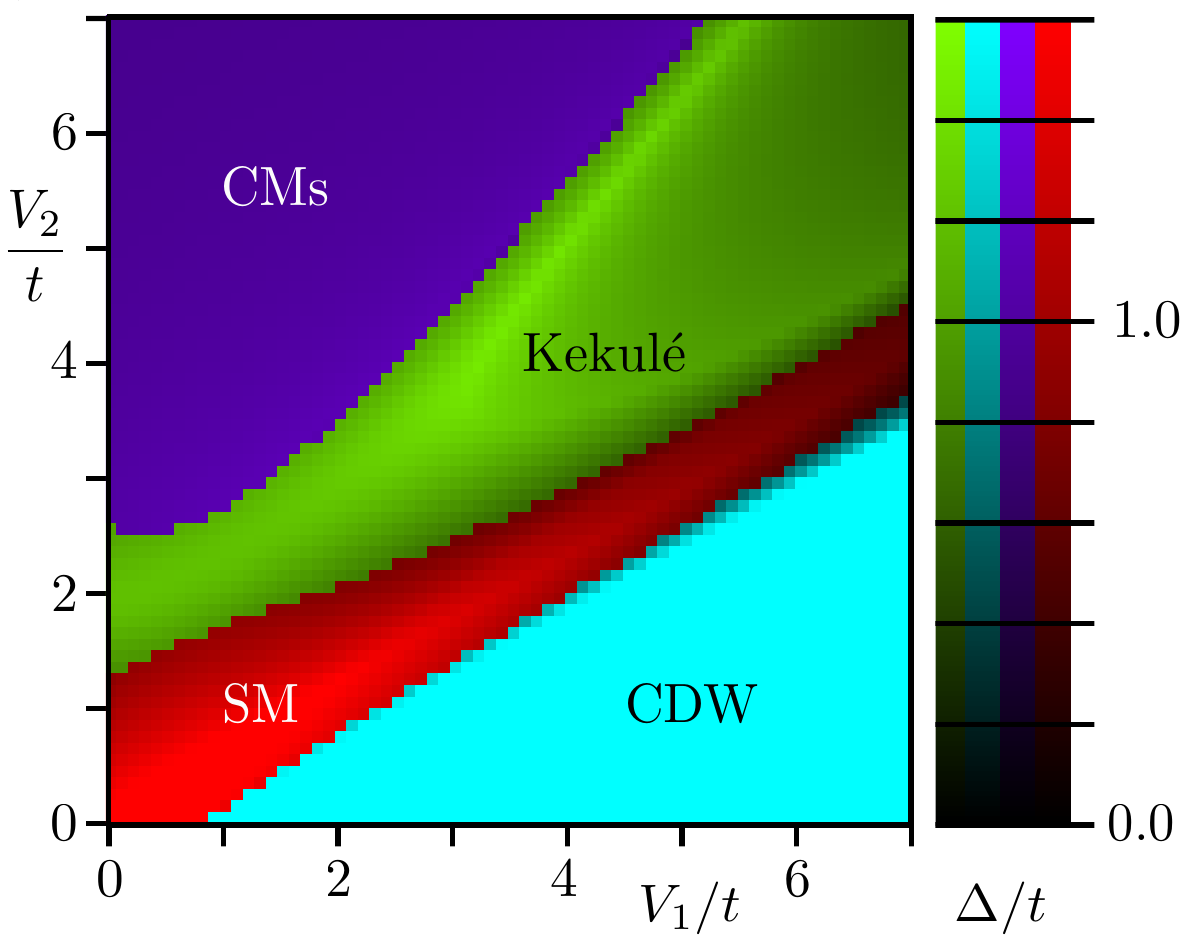}
\caption{(Color online) ED phase diagram for model (\ref{eq:Hspinless}) vs $(V_1/t,V_2/t)$ obtained using a cluster of $N=18$ sites. This is in rough agreement with most refined mean-field shown in Fig.~\ref{fig:MF3} with the notable difference that the topological QAH phase is not found. 
Reprinted figure with permission from Ref.~\onlinecite{Garcia2013} Copyright (2013) by the American Physical Society.
}
\label{fig:ED1}
\end{figure}

Regarding model (\ref{eq:Hspinless}), first ED results were provided in Ref.~\onlinecite{Garcia2013} based on numerical study of clusters with 18 and 24 sites. The phase diagram based on $N=18$ ED results is reproduced in Fig.~\ref{fig:ED1}. Note that this cluster being rather small, it has \emph{more} symmetries than the infinite one (translations and C$_{6v}$ point group symmetry), which can lead to artifacts. Based on that, the authors' major conclusions were that (i) there is a quite good agreement with the most refined mean-field~\cite{Grushin2013}, see Fig.~\ref{fig:MF3} except that the topological QAH is not realized; (ii) in particular, there is large portion of CM phase, with 18-fold degeneracy (which agrees with the strong coupling finding). 

\begin{figure}[!hb]
\includegraphics[width=\linewidth]{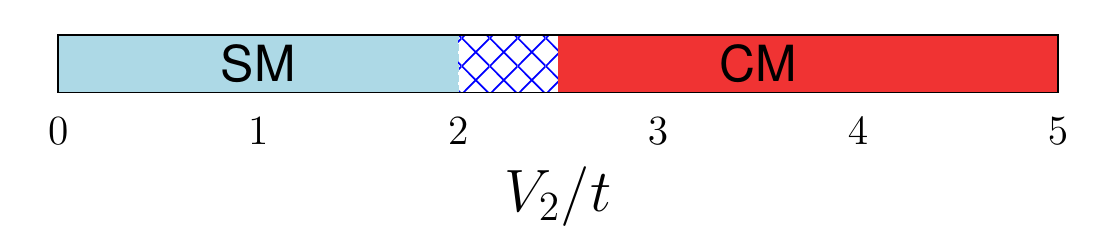}
\caption{(Color online) ED phase diagram for model (\ref{eq:Hspinless})  vs $V_2/t$ at fixed $V_1=0$ using a cluster with $N=24$ sites. The hatched region reflects the uncertainty regarding the direct transition between semimetallic (SM) and charge-modulated (CM) phase. 
Reprinted figure with permission from Ref.~\onlinecite{Daghofer2014} Copyright (2014) by the American Physical Society.
}
\label{fig:ED2}
\end{figure}

Soon after, another ED study based on clusters with $N=24$ and $N=30$ sites was published~\cite{Daghofer2014}. The authors have focused on the putative topological phase so that they have mostly considered $V_1=0$ case. Their phase diagram in Fig.~\ref{fig:ED2} shows that there is a direct transition between the semimetallic phase and the insulating CM phase at strong coupling, i.e. no intermediate Kekul\'e phase along this line contrary to Fig.~\ref{fig:ED1}. Note that there is a small difference regarding the nature of the CM phase, as compared to strong-coupling approach or previous ED reference, in the sense that they did not find charge imbalance between the sublattices. 

Then, an ED study was put forward using 18-site cluster but with open boundary conditions~\cite{Duric2014}. While it is rather small (there are less "bulk" sites than edge ones), a level crossing has been reported in the ground-state energy vs $V_2/t$ (at fixed $V_1=0$) corresponding to a change of parity with respect to inversion symmetry. This was taken as a \emph{positive} evidence for QAH and was supplemented by a variational Monte-Carlo approach using so-called entangled plaquette state (EPS) ansatz state. While the EPS approach is interesting, we believe that it can suffer from the same drawbacks as more standard mean-field approaches. Concerning ED with open boundary conditions, we have checked that the level crossing does not occur on the next $N=32$ cluster. 

\begin{figure}[!ht]
\includegraphics[width=\linewidth]{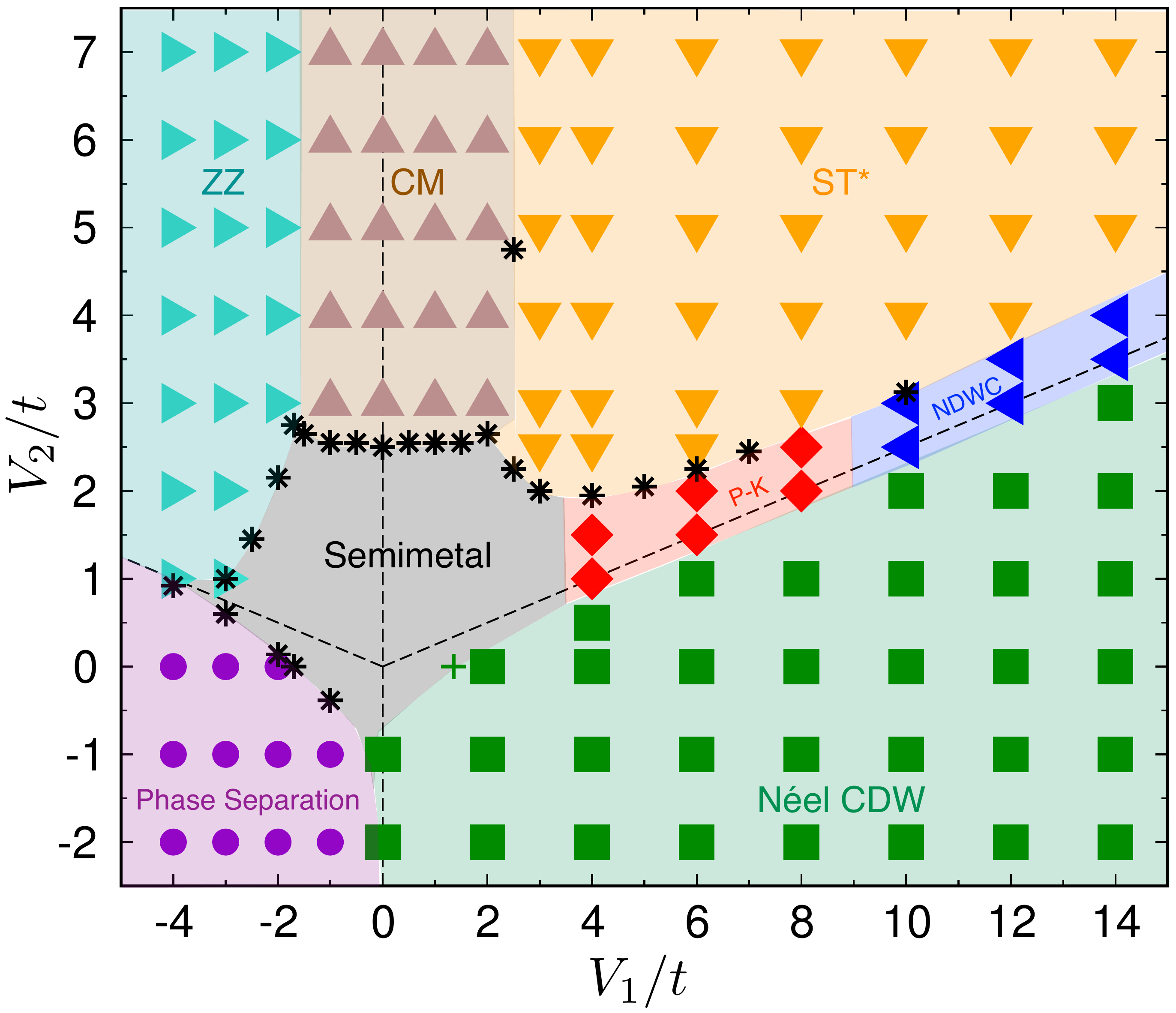}
\caption{(Color online) 
Phase diagram in the entire $(V_1/t,V_2/t)$ parameter space obtained using exact diagonalization techniques using clusters up to $N=42$ sites.  Dashed lines represent the classical transition lines, see Fig.~\ref{fig:IsingPhaseDiagram}. The semi-metal, which is the ground-state for non-interacting spinless fermions, has a finite extension in the phase diagram because of its vanishing density of states at the Fermi level. Several other phases can be stabilised for intermediate and/or large interactions: N\'eel CDW, plaquette/Kekul\'e (P-K), N\'eel domain wall crystal (NDWC), zigzag (ZZ) phase, and charge modulation (CM). The region (ST*) is degenerate at the semiclassical level, and it is presently unclear whether and how an order-by-disorder mechanism will lift the degeneracy.  Note also the large region of phase separation mostly in the attractive quadrant. 
Filled symbols correspond to numerical evidence (using level spectroscopy or measurements of correlations) obtained mostly on a $N=24$ cluster which contains the most important points in its Brillouin zone and features the full lattice point group symmetry of the honeycomb lattice. Star symbols denote likely first order transitions, witnessed by level crossings on the same cluster. Figure taken from Ref.~\onlinecite{Capponi2015}.
}
\label{fig:ED3}
\end{figure}

In our more recent paper~\cite{Capponi2015}, we have provided a more systematic ED approach by combining many more clusters up to $N=42$ sites. In particular, we have listed their geometric and symmetry properties as well as their Brillouin zone content. We refer to Ref.~\onlinecite{Capponi2015} for more details. Our main results are summarized in the phase diagram shown in Fig.~\ref{fig:ED3} and we will highlight some important features below.

First, let us mention that we have also investigated the attractive region ($V_1$ and/or $V_2$ negative) in order to make contact with our strong coupling approach. However, while the phase separation~\cite{Corboz2012} prevents some superconducting instabilities, it remains an open question to investigate if and where superconducting phases can be stabilized as suggested in the literature~\cite{Ryu2009,Roy2010,Roy2013,Roy2014,Jian2015,Kunst2015}.

Second, let us point out that the CM extension is much reduced as compared to Fig.~\ref{fig:ED1} since we do observe very sharp level crossings when increasing $V_1/t$ at fixed $V_2$. 

\begin{figure}[!htbp]
\includegraphics[width=\linewidth]{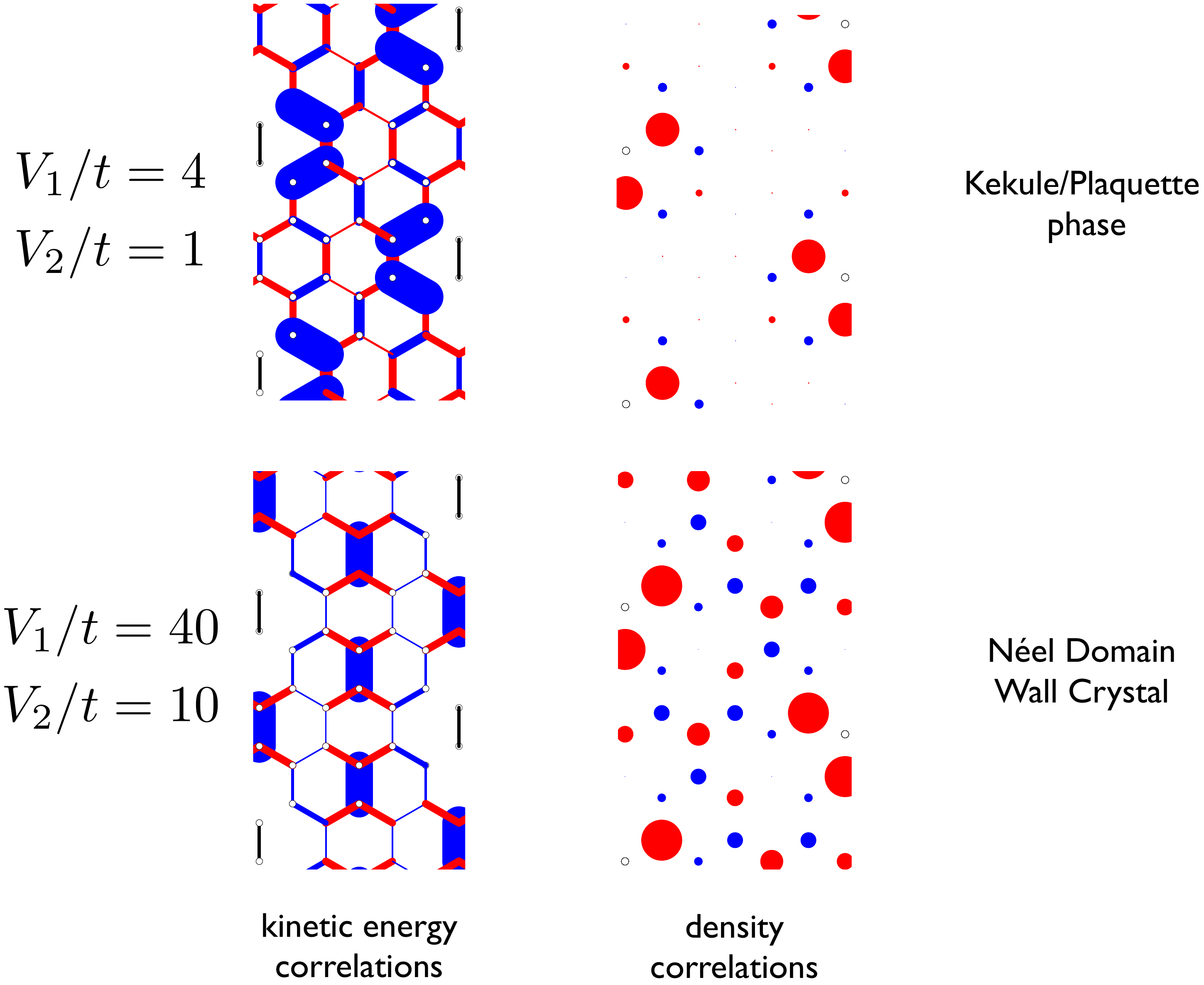}
\caption{(Color online) Kinetic energy and density (connected) correlations computed with ED on $N = 24$ cluster along the $V_1/V_2 = 4$ line, i.e. for $(V_1 /t, V_2 /t)$ respectively equal to $(4, 1)$ and $(40, 10)$. Blue and red correspond to positive/negative values. Reference bond is shown in black; reference site is an open circle. Periodic boundary conditions are used. 
Figure taken from Ref.~\onlinecite{Capponi2015}.
}
\label{fig:sketch}
\end{figure}

More importantly, and in order to make connection with our previous strong coupling analysis, we show in Fig.~\ref{fig:sketch} how the kinetic energy and density correlations drastically change along the $V_1/V_2=4$ line which was identified in the strong coupling approach. While there are indeed strong numerical evidence of a Kekul\'e pattern for intermediate interactions (as found in previous mean-field and ED approaches), it does not extend to the strong coupling regime where it is replaced by the NDWC phase, as expected see Fig.~\ref{fig:gs_eff_model_V1V2_4} and related discussion using the strong coupling approach.

Concerning the stability of the topological QAH phase, we have directly computed its order parameter using current-current correlations on a given sublattice. While the finite-size effects are highly non-trivial due to the large variety of cluster shapes (and hence different sets of $k$ points in their Brillouin zones), systematic extrapolations are compatible with a vanishing signal in the thermodynamic limit, i.e. the absence of topological phase in the phase diagram. 

\begin{figure*}[!ht]
\centering
\includegraphics[width=\linewidth]{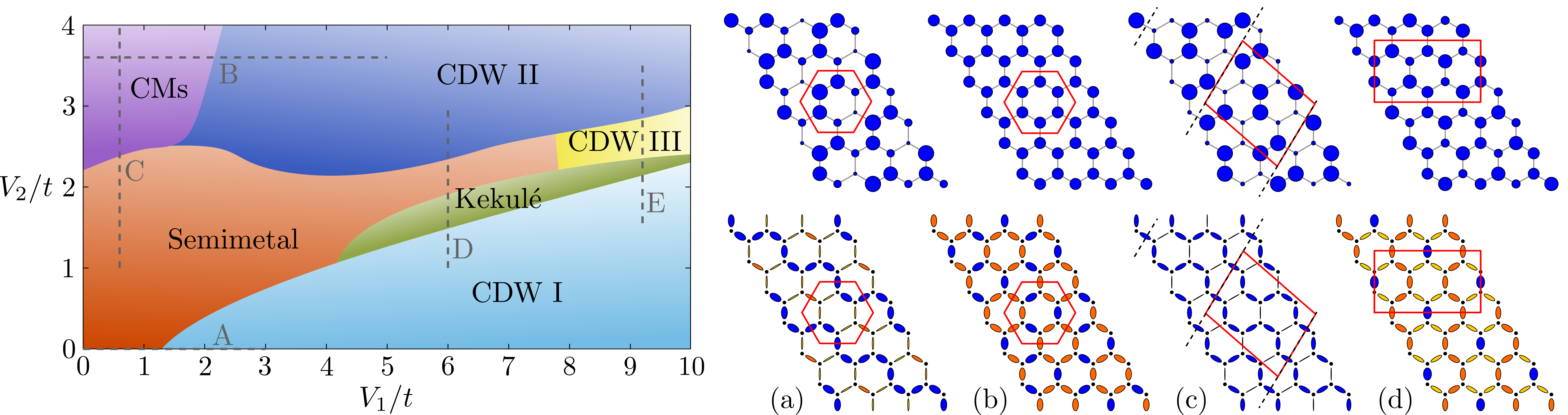}
\caption{(Color online) Left: Numerical phase diagram for repulsive $(V_1/t,V_2/t)$ interactions obtained with  DMRG calculations on a semi-infinite cylinder of width $L_y=12$ keeping up to $1,600$ states. Right: Various unit cells (in red polygons) that have been chosen as well as charge and bond strength patterns for various phases: (a) charge modulation (CM) with $V_1/t= 0.8$ and $V_2/t = 3.2$, (b) Kekul\'e phase with $V_1/t = 5.6$ and $V_2/t = 1.6$, (c) CDW II phase with $V_1/t = 5.6$ and $V_2/t = 3.2$, (d) CDW III phase (aka NDWC) with $V_1/t = 9.2$ and $V_2/t = 2.5$.
Reprinted figure with permission from Ref.~\onlinecite{Motruk2015} Copyright (2015) by the American Physical Society.
}
\label{fig:iDMRG}
\end{figure*}

\subsubsection{Density-Matrix Renormalization Group}
-- While DMRG was originally introduced for one-dimensional systems~\cite{White1992} where gapped phases of matter can be described using a finite-size matrix-product-state (MPS)~\cite{Schollwoeck2011}, it has  also become competitive in two dimensions in studying several frustrated quantum antiferromagnets for instance~\cite{Stoudenmire2012}.
In Ref.~\onlinecite{Motruk2015}, infinite DMRG algorithm has been applied to the spinless fermionic model~(\ref{eq:Hspinless}) using several possible unit cells and the numerical phase diagram is shown in Fig.~\ref{fig:iDMRG}.

Comparing with the numerical phase diagram obtained by ED in Fig.~\ref{fig:ED3}, we do observe a rather good semi-quantitative agreement, both on the nature of the extended phases as well as their locations. Most importantly, both studies do not support any region of topological QAH phase.

Note also that based on the entanglement entropy dependence on the number of kept states, the authors of 
Ref.~\onlinecite{Motruk2015} claim to have some indications about the nature of the various phase transitions. We refer to their publication for more details but we believe that, when possible, a deeper analysis of these phase transitions should be attempted, see below for instance. 

Last, let us point out that a finite sublattice charge imbalance was found in the CM phase  in agreement with our results~\cite{Capponi2015}. 

\subsubsection{Quantum Monte-Carlo}
-- Even in the simplest case ($V_2=0$) where a phase transition is expected at finite $V_1/t$ between the semi-metal and a N\'eel CDW state, stochastic QMC simulations were not possible  for a long time due to a severe sign-problem in the standard determinantal QMC agorithm~\cite{Scalapino1984,Gubernatis1985,Broecker2015}. As a consequence, the accuracy on the numerical critical value of the coupling $V_1^ c/t$, as well as the critical exponent of this continuous phase transitions were not well known until recently. 

I find this model rather important since it exemplifies several advances that have occurred in the QMC community, resulting in complete unbiased \emph{exact} results. First, so-called meron-cluster algorithm was used to solve the  sign problem for $V_1 \ge 2t$~\cite{Chandrasekharan1999}. More recently,  the sign problem has  been entirely solved for any $V_1>0$ by the continuous-time interaction expansion method~\cite{Rubtsov2005} using the Fermi bag idea~\cite{Huffman2014,Wang2014,Wang2015} and in the discrete-time method by using the Majorana fermion representation~\cite{Li2015}. It turns out that both solutions are possible thanks to a specific underlying Lie group structure of the determinantal QMC methods~\cite{Wang2015a}, which provides a useful guiding principle for sign-free QMC simulations.

\begin{figure}[!hb]
\includegraphics[width=\linewidth]{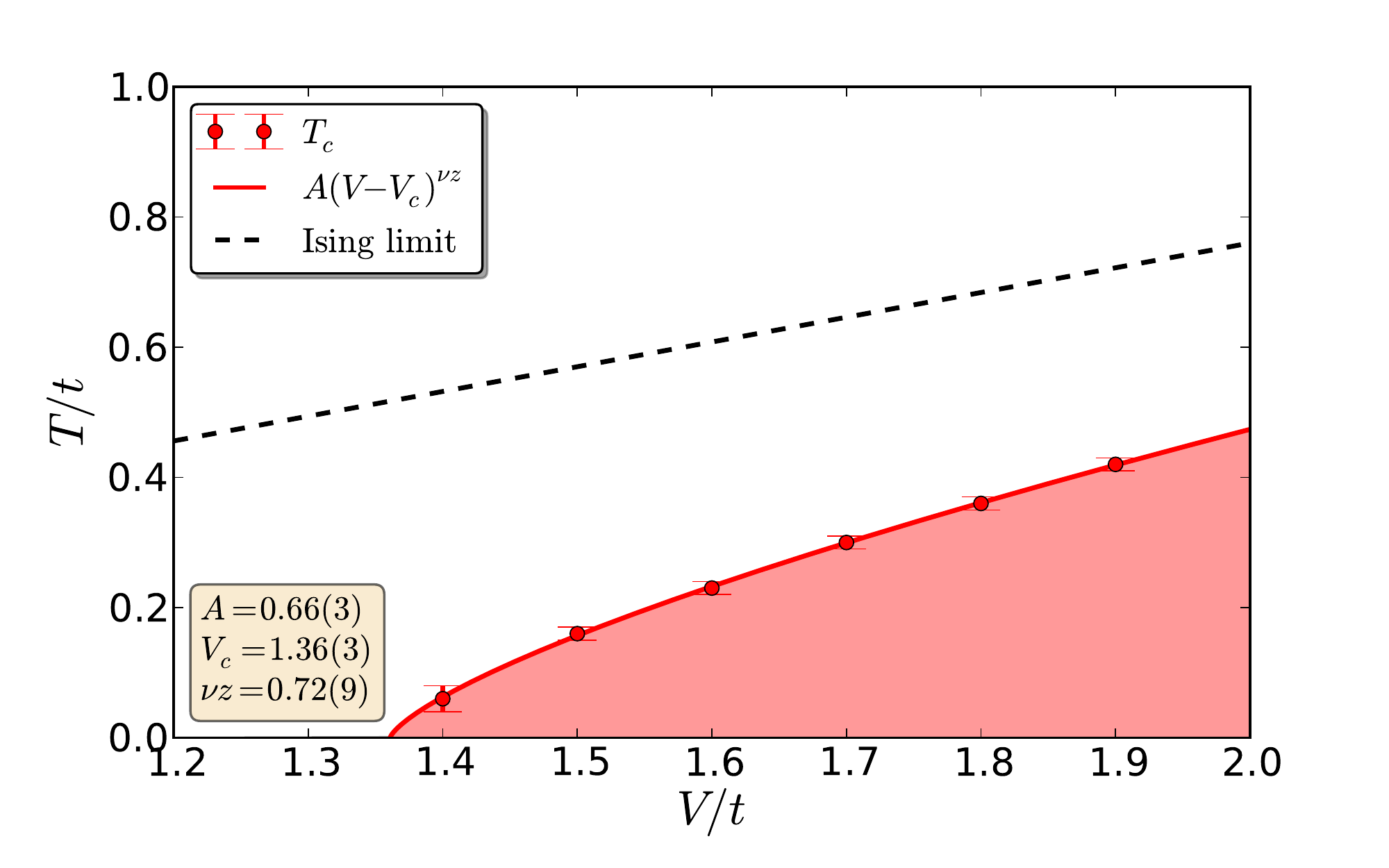}
\caption{(Color online) Phase diagram of the model (\ref{eq:Hspinless}) at $V_2=0$ as a function of $V_1/t$ and temperature on the honeycomb lattice. Shaded region corresponds to the N\'eel CDW I phase. The critical temperature $T_c$ approaches to zero at the quantum critical point between the CDW I and Dirac-semimetal (SM) state. The red solid line is a fit of the critical temperature to $T_c = A(V_1 - V_{1}^c)^{\nu z}$, leading to $V_1^c/t=1.36(3)$ and $\nu z=0.72(9)$. The dashed black line indicates the critical temperature in the Ising limit $T_c = 0.38V_1$, valid in  the strong-coupling limit $V_1 \gg t$. The quantum phase transition is in the Gross-Neveu-Yukawa with Z$_2$ order parameter universality class. 
Reprinted figure with permission from Ref.~\onlinecite{Wang2016} Copyright (2016) by the American Physical Society.
}
\label{fig:QMC}
\end{figure}

Recently, an even more efficient algorithm, based on stochastic series expansion (SSE)~\cite{Sandvik1991,Syljuasen2002} ideas, has been devised~\cite{Wang2016} allowing to study the repulsive $V_1$ ($V_2=0$) model and to obtain unbiased results on lattices up to $1,152$ sites at finite temperatures, thus revealing a very precise phase diagram in Fig.~\ref{fig:QMC} (see also Ref.~\onlinecite{Hesselmann2016}). In parallel, another algorithm using a Majorana representation has been used to study similar system sizes and results are in full agreement~\cite{Li2015a}. Moreover, by studying a different lattice model ($\pi$-flux square lattice) in the same universality class, it was confirmed that this quantum phase transition represents the 
 Gross-Neveu chiral-Ising universality class of two (two-component) Dirac fermions in 2+1D, and critical exponents are given~\cite{Li2015a} by $\eta = 0.45(2)$, $\nu= 0.77(3)$, and $\beta = 0.60(3)$.

Note that all the algorithms allowing to study this phase transition have a complexity ${\cal O}(\beta N^3)$, where $\beta$ is the inverse temperature and $N$ the number of sites, hence still not as competitive as existing algorithms for spin or bosonic models which behave as ${\cal O}(\beta N)$. For the sake of completeness, we point out that consistent results have been obtained using infinite projected entangled-pair states (iPEPS) algorithm~\cite{Wang2014} which aims at optimizing the ground-state wavefunction in a tensor-network representation. The advantage is that iPEPS could be used for any parameters, which looks promising to characterize the various phase transitions for instance.

Let us also mention an interesting idea to tackle with the sign-problem in QMC simulations: while standard measurements are not possible, it could be useful to use machine-learning algorithm in order to compare the wavefunction to known ground-states for certain parameters (non-interacting case or simple CDW for instance) so that one could in principle be able to detect some quantum phase transitions~\cite{Broecker2016}.

Despite these impressive developments, it remains impossible to study the finite $V_2>0$ case without sign problem, so that the complete phase diagram cannot be obtained with QMC techniques.


\begin{figure}[!hb]
\includegraphics[width=\linewidth]{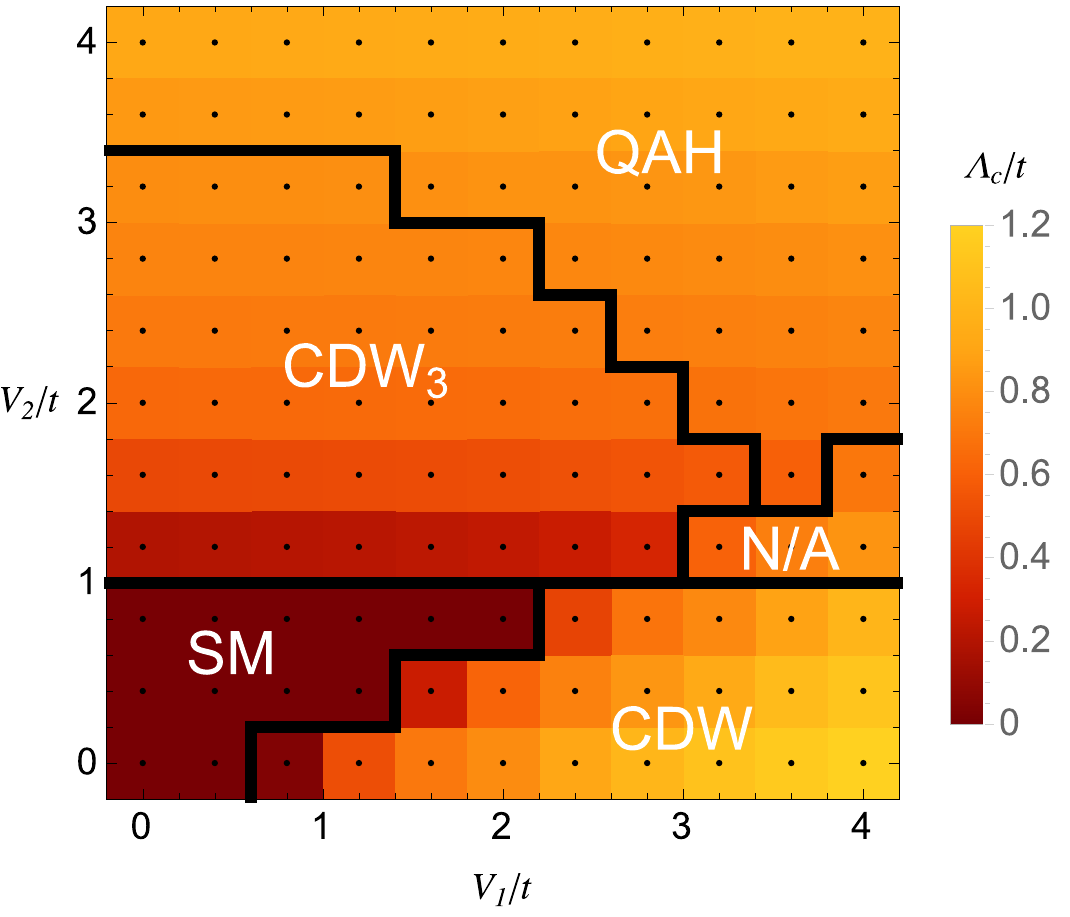}
\caption{(Color online) Phase diagram of the model (\ref{eq:Hspinless}) obtained using fRG technique. CDW$_3$ denotes the CM phase, which occurs when increasing $V_2$ at $V_1=0$. For very large interactions, the QAH instability emerges in the phase diagram but it is believed to be an artifact due to the breakdown of the weak coupling approximation. In the region marked with N/A,  the leading instability could not be determined.
Reprinted figure with permission from Ref.~\onlinecite{Scherer2015} Copyright (2015) by the American Physical Society.
}
\label{fig:fRG}
\end{figure}

\subsubsection{Functional renormalization group approach}
-- This technique aims at determining the leading instability in the weak-coupling regime of a Fermi gas subject to some interactions. It is rather technical and requires several approximations to patch the Fermi surface for instance. The most recent phase diagram for the spinless case that we consider in this review is reproduced in Fig.~\ref{fig:fRG} from Ref.~\onlinecite{Scherer2015}. As found in most previous numerical approaches, there is a direct transition from the semimetal to the CM phase when increasing $V_2$, hence no QAH phase. It appears at larger interaction but the authors believe that this may be an artifact since they are using a weak-coupling approach. Concerning the simpler case with interaction $V_1$ only, the critical value is found to be $0.6 t$ which is quite far from the exact reported QMC value ($1.36 t$, see Fig.~\ref{fig:QMC}).

As a side remark, in the spinful case, recent fRG analysis have also concluded to the absence of topological (QSH) phase~\cite{Volpez2016,SanchezdelaPena2016}. 

However, we would like to remind the reader that in the seminal paper by Raghu {\it et al.}~\cite{Raghu2008}, some similar fRG results were presented and pointed to the stability of QSH phase, quite opposite to these more recent publications using a higher momentum-space resolution.

\section{Conclusions and outlook}

We have reviewed recent works on correlated spinless fermions on the honeycomb lattice at half-filling. 
While being a rather simple model, correlations are responsible for a large variety of competing phases. In particular, mean-field study has pointed out the exciting possibility to stabilize a Chern insulating phase with topological character. This has been a numerical challenge for some years and we have tried to summarize the most relevant results. For sure, numerical data have substantially modified the mean-field phase diagram: several kind of charge orderings have been demonstrated and the topological phase has not been confirmed (in most studies). 
While the absence of QAH sounds like a negative result, there are encouraging signals that it is a competing nearby phase, presumably close in energy. It is possible that a slight modification of the microscopic model could favor it: for instance, longer distance RKKY interactions have been proposed to stabilize the QAH phase at the mean-field level~\cite{Liu2016}. Similarly, on the kagome lattice at 1/3 filling, a recent DMRG study has shown a small region of QAH phase including up to third-neighbor density interactions~ \cite{Zhu2016}. It could be also interesting to investigate the square lattice with a $\pi$-flux which has a similar band structure and where mean-field also predicts emergence of a topological phase, although ED numerical study could not detect it~\cite{Jia2013}.

On the other hand, showing the absence of QAH phase in this microscopic model has required to improve various numerical and analytical techniques. In particular, we believe that understanding the strong coupling regime in these systems might be a useful strategy to complement other weak-coupling techniques. We have demonstrated this by determining new charge modulations that occur in this limit, and that were confirmed numerically after. This could be a useful approach to other strongly correlated systems.

Last but not least, it will be interesting to investigate the case away from half-filling where many other phases have been proposed~\cite{Castro2011,Grushin2013}.

The author is grateful to 
A. L\"auchli for 
collaboration on this work and to 
M. Daghofer,
A. Grushin,
M. Hohenadler,
V. Juri\ifmmode \check{c}\else \v{c}\fi{}i\ifmmode \acute{c}\else \'{c}\fi{},
J. Motruk, 
F. Pollmann,
B. Roy,  
D. Scherer, 
S. Trebst, 
L. Wang, and
S. Wessel
for valuable discussions and comments.
\section*{References}

\bibliographystyle{prsty}

\end{document}